\documentclass[conference]{IEEEtran}
\IEEEoverridecommandlockouts
\usepackage{cite}
\usepackage{amsmath,amssymb,amsfonts}
\usepackage{graphicx}
\usepackage{textcomp}
\def\BibTeX{{\rm B\kern-.05em{\sc i\kern-.025em b}\kern-.08em
    T\kern-.1667em\lower.7ex\hbox{E}\kern-.125emX}}

\usepackage[dvipsnames]{xcolor}
\usepackage{bibentry}
\usepackage[most]{tcolorbox}
\newcommand{\blue}[1]{\textcolor{blue}{#1}}
\usepackage{pgfplots}
\usepackage{multicol}
\usepackage{subfigure}
\usepackage{tikz}
\usepackage[ruled,vlined,linesnumbered]{algorithm2e}
\usepackage{booktabs}
\usepackage{multirow}   
\usepackage{setspace}
\usepackage{hyperref}
\usepackage{enumitem}
\usepackage{makecell}
\usepackage{soul}

\newcommand{\RNum}[1]{\uppercase\expandafter{\romannumeral #1\relax}}
\pgfplotsset{compat=1.18}

\begin{document}

\title{MRP-LLM: Multitask Reflective Large Language Models for Privacy-Preserving  Next POI Recommendation}


\author{
    Ziqing Wu$^{1}$, 
    Zhu Sun$^{2, *}$\thanks{*Corresponding author}, 
    Dongxia Wang$^{3}$, 
    Lu Zhang$^{4}$, 
    Jie Zhang$^{1}$, 
    Yew Soon Ong$^{1}$\\
    \IEEEauthorblockA{
        $^{1}$Nanyang Technological University, Singapore, Singapore \\
        $^{2}$Singapore University of Technology and Design, Singapore, Singapore \\
        $^{3}$Zhejiang University, Hangzhou, China \\
        $^{4}$Chengdu University of Information Technology, Chengdu, China \\
        ziqing002@e.ntu.edu.sg, sunzhuntu@gmail.com, dxwang@zju.edu.cn, \\
        zhang\_lu010@outlook.com, 
        zhangj@ntu.edu.sg, asysong@ntu.edu.sg
    }
}


\maketitle

\begin{abstract}
Large language models (LLMs) have shown promising potential for next Point-of-Interest (POI) recommendation. However, existing methods only perform direct zero-shot prompting, leading to ineffective extraction of user preferences, insufficient injection of collaborative signals, and a lack of user privacy protection. As such, we propose a novel \textbf{\underline{M}}ultitask \textbf{\underline{R}}eflective \textbf{\underline{L}}arge \textbf{\underline{L}}anguage \textbf{\underline{M}}odel for \textbf{\underline{P}}rivacy-preserving  Next POI Recommendation (MRP-LLM), aiming to exploit LLMs for better next POI recommendation while preserving user privacy. Specifically, the \textit{Multitask Reflective Preference Extraction Module} first utilizes LLMs to distill each user's fine-grained (i.e., categorical, temporal, and spatial) preferences into a knowledge base (KB). 
The \textit{Neighbor Preference Retrieval Module} retrieves and summarizes the preferences of similar users from the KB to obtain collaborative signals. Subsequently, aggregating the user’s preferences with those of similar users, the \textit{Multitask Next POI Recommendation Module} generates the next POI recommendations via multitask prompting.
Meanwhile, during data collection, a \textit{Privacy Transmission Module} is specifically devised to preserve sensitive POI data. Extensive experiments on three real-world datasets demonstrate the efficacy of our proposed MRP-LLM in providing more accurate next POI recommendations with user privacy preserved.
\end{abstract}

\begin{IEEEkeywords}
next poi recommendation, large language models, privacy preservation
\end{IEEEkeywords}

\section{Introduction}\label{sec:introduction}
Next Point-of-Interest (POI) recommendation has been widely utilized on location-based social networks (LBSNs) such as Yelp\footnote{\url{https://www.yelp.com/}} and Foursquare\footnote{\url{https://foursquare.com/}}. Its goal is to learn users' check-in preferences and suggest relevant locations for their next visit, thereby reducing the information overload and enhancing user experiences~\cite{sanchez-et-al:point}.

Conventional next POI recommender systems (RSs) usually train a domain-specific model based on users' check-in sequences to identify users' preferences on locations. In pursuit of higher accuracy, next POI RSs have evolved from Markov-chain-based methods~\cite{rendle:2010factorizing} and RNN-based methods~\cite{liu:2016predict} to transformer-based methods~\cite{luo:2021stan} and graph-based methods~\cite{lim:2022hierarchical}. 
Despite the success, they still face challenges including data sparsity and cold-start users.
Fortunately, recent advancements in large language models (LLMs) such as Chat-GPT\footnote{\url{https://chat.openai.com}} have presented new opportunities to overcome the limitations and further enhance next POI RSs.  
First, LLMs have demonstrated the ability to transfer knowledge, such as spatial and temporal information, from the training corpus of other domains~\cite{DBLP:conf/iclr/GurneeT24}, which could facilitate users' POI preferences learning even with sparse user data.
Second, in-context learning (ICL) has demonstrated promising potential in delivering effective next-item recommendations, enabling robust learning even for cold start users~\cite{wang2023drdt}.
Third, due to their natural language reasoning capabilities, LLMs can provide explainable hints for their recommendations, thereby improving the interpretability of the recommendations as well as users' trust in the system.

Despite their remarkable achievements across various recommendation domains~\cite{sun2023dynamic,sun2024large}, LLMs for next POI recommendation remain a relatively unexplored area. Existing work attempts to perform zero-shot prompting 
based on users' own check-in history~\cite{wang2023would}. 
However, such an approach exhibits several obvious shortcomings. 
\begin{enumerate}
    \item \textit{Ineffective extraction of user preferences.} Successful POI RSs usually focus on accurately capturing users' POI-related preferences~\cite{zhang:2021interactive}.
    Instead, LLMs-based next POI RSs lack a clear reasoning of the impact of fine-grained user preferences on their decisions. 
    \textcolor{black}{Thus, they have shown weaknesses in understanding the spatial-temporal information of user check-ins and mining user transition preferences, which can undermine the accuracy~\cite{feng2024move}.} 
    \item \textit{Insufficient injection of collaborative signals.} Conventional next POI RSs use collaborative signals by analyzing behaviors of similar users to obtain global behavioral patterns and enhance both accuracy and generalizability. In contrast, existing LLM-based methods rely exclusively on the user's own check-in history, resulting in a limited view that does not incorporate insights from broader datasets.
    \item \textit{Lacking of privacy protection.} The existing LLM-based methods directly expose user check-in history to LLMs, which may contain sensitive information such as users' whereabouts, home addresses, and behavioral habits~\cite{ackermann-et-al:2022willingness}. The potential sensitive information leakage could erode user trust in next POI RSs.
\end{enumerate}

\begin{figure}[t]
    \centering
    \includegraphics[width=\linewidth]{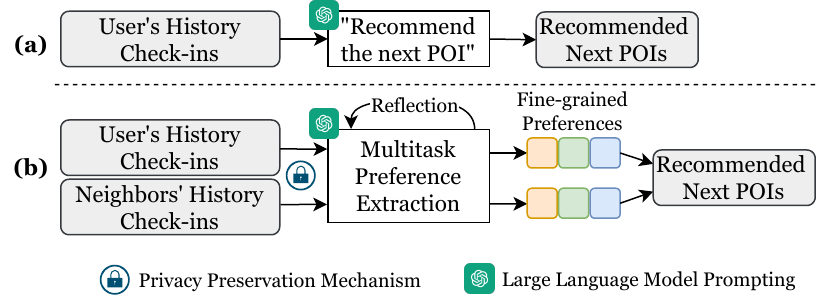}
    \caption{Our proposed method shown in (b), which makes recommendations based on more diverse and fine-grained user preferences, compared to existing direct zero-shot methods shown in (a).}
    \label{fig:2}
\end{figure}

%
%

Therefore, we propose a novel \textbf{M}ultitask \textbf{R}eflective \textbf{L}arge \textbf{L}anguage \textbf{M}odel for \textbf{P}rivacy-preserving Next POI Recommendation (MRP-LLM) to address the identified issues.
As illustrated in Figure~\ref{fig:2}, 
MRP-LLM incorporates the user's and her neighbors' fine-grained preferences extracted from multitask prompting and self-reflection mechanisms to improve recommendation accuracy. Moreover, sensitive check-in-related inputs are protected when uploading to the RS.

Specifically, to effectively extract users' 
behavioral patterns, a \textit{Multitask Reflective Preference Extraction Module} first distills each user's categorical, temporal, and spatial preferences on POIs into a fine-grained preference knowledge base (KB). 
\textcolor{black}{The \textit{Neighbor Preference Retrieval Module} then identifies users' neighbors based on their historical distribution and social relationship, and retrieves and summarizes their preferences from the KB to inject collaborative signals.}
\textcolor{black}{Subsequently, the \textit{Multitask Next POI Recommendation Module} leverages the user's preferences with those of her neighbors to facilitate the next POI recommendation based on her check-in history.}
%
Moreover, to mitigate sensitive information leakage, \textcolor{black}{check-in history and social relationships}
are retained locally. A \textit{Privacy Transmission Module} is specially devised to leverage differential privacy to safeguard each type of user data uploaded to the RS.
As such, MRP-LLM not only exploits the LLMs to enhance recommendation accuracy but also provides robust protection for user data privacy.

In summary, the contributions of this study lie three-fold.
\begin{enumerate}[leftmargin=15pt]
    \item We propose a novel multitask reflective LLM-based next POI RS (MRP-LLM), which could enhance recommendation accuracy by reflectively extracting fine-grained user preferences and integrating collaborative signals from neighbors into the LLM recommendation process.
    \item \textcolor{black}{We design a novel differential privacy mechanism to comprehensively protect the sensitive user data including users' check-in history and social relationships}
    for more secure LLM-based next POI recommendations.
    \item We conduct extensive experiments on three real-world datasets to verify the superiority of MRP-LLM over state-of-the-art methods (SOTAs). \textcolor{black}{Specifically, when privacy protection is relaxed, it gains an average lift of 8.4\% and 7.0\% in ACC and MRR; with full-scale protection on users' sensitive data, it still achieves performance comparable to other LLM-based next POI RSs, with a slight drop of 1.3\% in ACC, and a lift of 0.8\% in MRR.}
\end{enumerate}

\section{Related Work}\label{sec:related}

Although the application of LLMs in next POI recommendation is less explored, both fields have been extensively studied independently. Additionally, privacy preservation for both LLMs and RSs has received considerable attention in recent research.
This section reviews related work in four key areas to our research, the next POI recommendation, the LLM-based recommendation, the privacy-preserving approaches for conventional RSs, and the privacy-preservation techniques specifically designed for LLMs.

\subsection{Next POI Recommendation}

Early studies on next POI recommendation such as FPMC-LR~\cite{cheng2013you} utilize Markov chain models to learn users' consecutive check-in behaviors~\cite{rendle:2010factorizing}. 
%
With the evolution of recurrent neural networks (RNNs)~\cite{sherstinsky2020fundamentals}, researchers start to model the sequential check-in by incorporating POI-specific spatial and temporal signals into RNN structures. For example, ST-RNN~\cite{liu:2016predict} introduces time- and distance-specific transition matrices to enhance RNNs' capability of capturing spatiotemporal transition signals; STGN~\cite{zhao:2020go} introduces time and distance gates
into the structure of long short-term memory network(LSTM)~\cite{hochreiter1997long}.

Later methods utilize the advanced capabilities of attention mechanisms and transformer architectures~\cite{vaswani2017attention}. Methods such as Deepmove~\cite{feng:2018deepmove} and ATST-LSTM~\cite{huang:2019attention} fuse the attention network into RNN and LSTM model respectively to capture periodicity and spatio-temporal context; STAN~\cite{luo:2021stan} and CFPRec~\cite{zhang2022next} both achieve remarkable recommendation accuracy by leveraging the bi-layer attention and bidirectional transformer architecture.

Recent works further explore the integration of graph-based models to further enhance performance. For example, HMT-GRN~\cite{lim:2022hierarchical} models POI relations with a graph recurrent network on both temporal and spatial POI-POI graphs;  STHGCN~\cite{yan2023spatio} learns the spatial semantic relationships between check-ins with graph convolutional networks (GCN); DCHL~\cite{lai2024disentangled} adopt adjusted hypergraph convolutional networks on users' check-in graph to learn the evolution of users' multi-aspect preferences.
Nevertheless, all conventional next POI recommenders face challenges of data sparsity and lack of interpretability. 

\subsection{LLM-based Recommendation}

\subsubsection{General Recommendation}
Thanks to the extensive knowledge and powerful reasoning capability, LLMs have gained much attention in the field of recommendation. 
One popular branch is to leverage the zero-shot ability of LLMs by directly prompting LLMs with recommendation-specific instructions. For instance, Liu et al.~\cite{liu2023chatgpt} first investigate the capability of ChatGPT to transfer knowledge from other domains to perform five different recommendation tasks such as rating prediction, recommendation generation, and explanation. 

To achieve better performance, later methods attempt to obtain more effective prompts via various means. For example, Wang et al.~\cite{wang2023zero} \textcolor{black}{apply Chain-of-Thought (CoT) technique~\cite{wei2022chain}} with a three-stage prompting to extract user preferences on movies step-by-step; 
\textcolor{black}{Sun et al.~\cite{sun2023chatgpt} proposes a sliding window prompting strategy that repeatedly ranks candidates within a sliding window to obtain overall rankings.}
%
Other methods utilize LLMs as a supplement to conventional RSs. 
For example, KAR~\cite{xi2024towards} encodes the reasoning and factual knowledge provided by LLMs as vectors to enrich the user and item representations in RSs; LLMRec~\cite{wei2024llmrec} utilize LLMs to augment user and item side information in the GCN recommender; and GaCLLM~\cite{du2024large} utilize LLMs as means for propagation and aggregation in graph-based RSs.

Lastly, some works attempt to fine-tune the LLMs on the recommendation tasks. For example, GPTRec~\cite{petrov2023generative} fine-tunes a GPT2 model for generative recommendation tasks; P5~\cite{geng2022recommendation} fine-tunes the T5 model concurrently on five recommendation tasks to achieve zero-shot generalization; Tallrec~\cite{bao2023tallrec} uses LoRA to pre-train LLaMA based on Alpaca and a recommendation-specific dataset; Once~\cite{liu2024once} proposes to use prompting for close-source LLMs such as GPT3 together with the fine-tuned open-source LLMs such as LLaMA to boost the recommendation performance. 

\subsubsection{Sequential Recommendation}
LLMs demonstrated potential in processing sequential user-item interaction history and performing next-item recommendations. 
\textcolor{black}{Using user purchase history sequences, Gao et al.~\cite{gao2023chat} employ zero-shot prompting to directly instruct LLMs to recommend the next item.
Liu et al.~\cite{liu2023chatgpt} and Hou et al.~\cite{hou2024large} verify that few-shot prompting may help LLMs generate more accurate next item recommendations.
Furthermore, later works attempt to leverage more advanced techniques to boost ICL performance. For example, DRDT~\cite{wang2023drdt} and LLM4ISR~\cite{sun2024large} have demonstrated that incorporating a self-reflection mechanism could help LLMs capture users' preference evolution or optimize prompts for sequential recommendation tasks; Re2LLM~\cite{wang2024re2llm} train a lightweight reinforcement learning model to retrieve hints generated by LLMs from external knowledge bases to enhance task instructions.
%
}
%
%
Recent attempts have also begun to explore using LLMs for next POI RSs. LLMMob~\cite{wang2023would} employed zero-shot prompting for next POI recommendation based on individual users' check-in data and time slot; LLMMove~\cite{feng2024move} further incorporates the geographical information of the POIs. However, they are unable to effectively extract fine-grained user preferences. Another line is to fine-tune LLMs on the next POI recommendation task but works like LLM4POI~\cite{li2024large} demand substantial training data and computational resources.

\subsection{Privacy-Preservation for Recommendation}

User privacy has been recognized as a prominent problem and widely discussed for conventional RSs. In order to protect users' interaction history and other side information, recent works mainly leverage cryptographic techniques and perturbations to prevent privacy leakage.
Cryptographic techniques ensure sensitive user data remains secure and inaccessible to unauthorized parties, including the RS itself. For example, Zhou et al.~\cite{zhou2023lightweight} propose a lightweight fully homomorphic data encapsulation mechanism in a distributed RS and ensure that both user rating data and recommendation results are kept private against the RS; PCAPR~\cite{sun2024towards} uses symmetric homomorphic encryption to detect users' physical distance and social distance privately for more accurate recommendation. 
The perturbation method achieves an individual's privacy by adding noise to the data or computations to mask individual contributions and achieving differential privacy~\cite{dwork:2014algorithmic}.
Compared to cryptographic techniques, differential privacy perturbation typically requires simpler communication protocols and is better suited for deep learning-based RSs with complex architectures and computations.

Specifically for POI RSs, the protection usually focuses on users' raw check-in data with data isolation~\cite{long2023model}, encryption~\cite{perifanis2023fedpoirec} or perturbation techniques~\cite{dong2022ranking}. 
Some work further protects users' preferences and friendship relations~\cite{long2023decentralized}.
Another major privacy challenge is to protect the spatial information of check-ins. To retain the spatial closeness without revealing the actual position of the POI, methods have been proposed to guarantee differential privacy~\cite{lu2008pad,huo2021privacy}.

\subsection{Privacy-Preservation for LLMs}
Although LLMs have shown substantial potential across various research domains, their application has also introduced new privacy concerns. In particular, LLMs may first be susceptible to passive privacy leakage, i.e., LLMs may expose sensitive information in user prompts or training corpus~\cite{kshetri2023cybercrime}. What is more, LLMs may also be vulnerable to active privacy attacks such as model inversion attacks~\cite{zhang2022text}, and membership inference attacks~\cite{mireshghallah2022quantifying}, which can reveal private texts or confirm the existence of specific training samples. 
Thus, some works have been proposed to tackle the emerging privacy risks of LLMs at the fine-tuning and inference stage. For fine-tuning LLMs, works such as EW-Tune~\cite{behnia2022ew} and JFT~\cite{shi2022just} apply differential privacy to the tuning data. Other works such as FedBPT~\cite{sunfedbpt} fine-tune LLMs with federated learning to avoid raw data sharing. For LLM inference, primitive methods detect and eliminate sensitive inputs in user prompts~\cite{kim2024propile}. Cryptographic methods including fully homomorphic encryption~\cite{lu2023bumblebee} and multi-party computation~\cite{dong2023puma} allow private user input to Transformer-based models. Differential privacy could also be applied to protect privacy text generation. For example, Mai et al.~\cite{mai2023split} apply local differential privacy on LLM output embeddings to prevent inference attacks. 

Specifically for privacy-preserving LLM-based RSs, some methods focus on privacy protection for centralized service providers, assuming that these providers are completely honest~\cite{shi2022selective}.
Other methods, such as RAPT~\cite{li2023privacy}, attempt to perturb sensitive keywords in user prompts before submitting them to LLMs, thereby obscuring sensitive inputs within the prompts.
So far, limited studies have examined privacy solutions for LLM-based RSs. A recent attempt leverages the differential back-propagation techniques to fine-tune a privacy-preserving LLM, which requires excessive computation resources~\cite{carranza2023privacy}.
Although privacy protection has gained attention for general LLMs, privacy protection in LLM-based RSs, especially the next POI recommendation tasks, requires further exploration.

\section{Methodology}\label{sec:method}

\tcbset{
    enhanced,
    colframe=BrickRed, 
    colback=white, 
    coltitle=white, 
    fonttitle=\bfseries,
    title={},
    boxrule=0.5mm, 
    rounded corners,
    width=0.95\linewidth, 
    arc=1mm, 
    before=\vspace{0.5em}, 
    after=\vspace{0.5em}, 
    boxsep=0pt, 
    left skip=0pt, 
    right skip=5pt,
    top=3mm,                
    bottom=3mm,             
    left=5mm,               
    right=5mm,              
}
\def\thesubsubsectiondis{\unskip\arabic{subsubsection})}

\begin{figure*}[t]
    \centering
\includegraphics[width=0.9\textwidth]{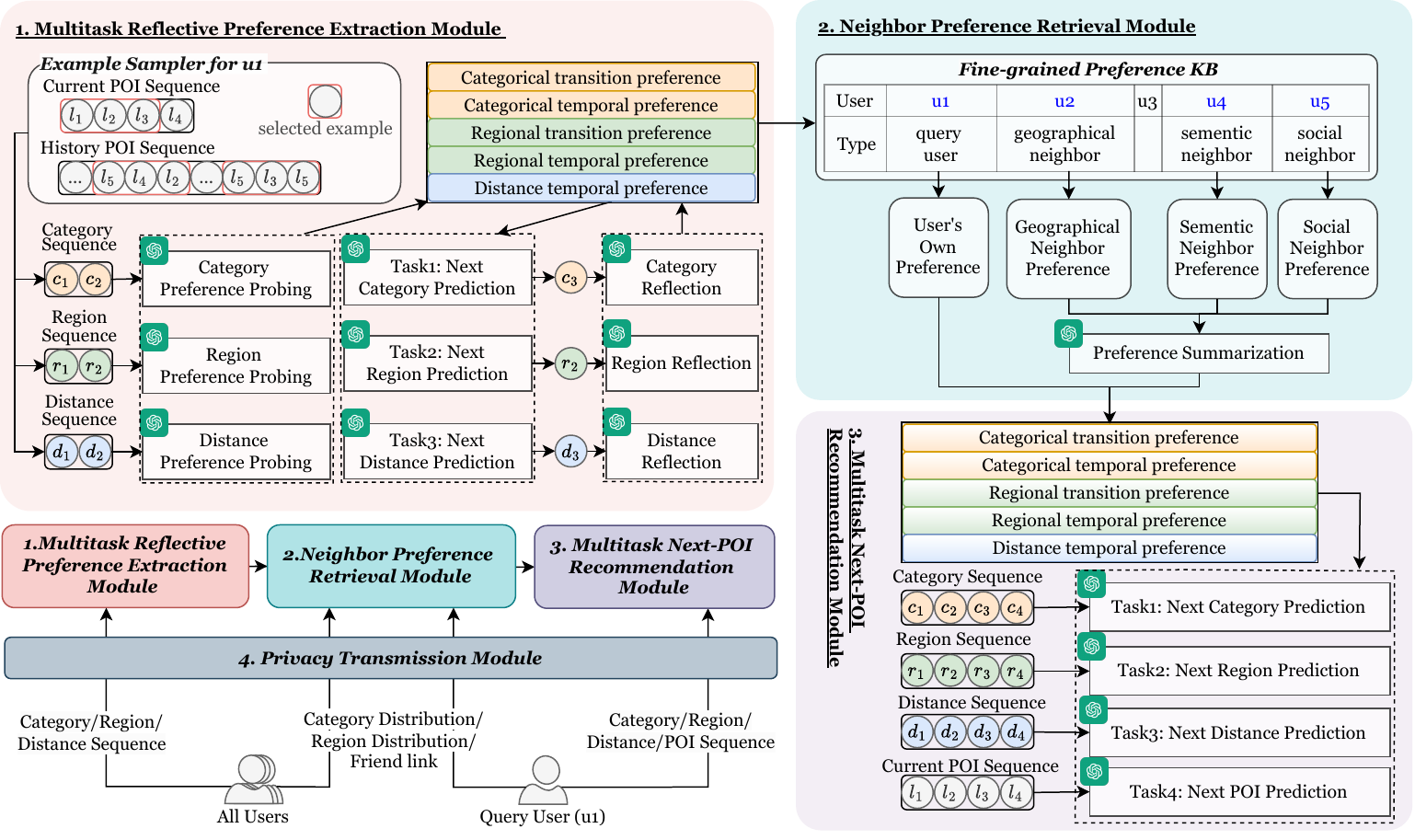}
    \caption{The architecture of MRP-LLM. Taking $u_1$ as the recommendation query user.} 
    \label{fig:1}
\end{figure*}

\subsection{Notations and Problem Statement}
\noindent\textbf{Notations}.
Let $\mathcal{U}=\{u_1, \allowbreak u_2, \allowbreak ..., \allowbreak u_{|\mathcal{U}|}\}$, $\mathcal{L}=\{l_1, l_2,...l_{|\mathcal{L}|}\}$, $\mathcal{C}=\{c_1, c_2,...c_{|\mathcal{C}|}\}$ and $\mathcal{R}=\{r_1, r_2,...r_{|\mathcal{R}|}\}$ denote the sets of users, POIs, POI categories, and POI regions respectively\footnote{In our study, the terms `POI' and `Location' are exchangeable.}. \textcolor{black}{Users' social relations are represented by matrix $\mathbf{S}$.}
A check-in record $(u, l, t, r, c, d)$ denotes user $u$ visited POI $l$ at time $t$, where $l$ located in region $r$ is characterized by category $c$ and is $d$ km away from the user's previous check-in.
For each user, we chronologically order her check-in records to form check-in sequences. The most recently visited POI sequence is the \textit{current POI sequence}, $\mathcal{L}_u^{cur}$, whereas the POIs visited previously form the \textit{history POI sequence}, $\mathcal{L}_u^{hist}$. Both current and history POI sequences comprise a list of POIs and check-in time, i.e., $\mathcal{L}_u=\{(l_{t_1}, t_1), (l_{t_2}, t_2), ..., (l_{t_k}, t_k)\}$. 
The corresponding category, region, and distance sequences are 
$\mathcal{C}_u=\{(c_{t_1}, t_1), \allowbreak (c_{t_2}, t_2), \allowbreak ..., \allowbreak (c_{t_k}, t_k)\}$, 
$\mathcal{R}_u=\{(r_{t_1}, t_1), \allowbreak (r_{t_2}, t_2), \allowbreak ..., \allowbreak (r_{t_k}, t_k)\}$, 
and $\mathcal{D}_u=\{(d_{t_1}, t_1), \allowbreak (d_{t_2}, t_2), \allowbreak ..., \allowbreak (d_{t_k}, t_k)\}$, respectively. The important notations utilized in our paper are summarized in Table~\ref{tab:notation}.

\smallskip\noindent\textbf{Problem Statement}.
\textcolor{black}{The task of MRP-LLM is to recommend the next POI $l_{t_{k+1}}$ given users' current sequences $\mathcal{L}_u^{cur}, \mathcal{C}_u^{cur}, \mathcal{R}_u^{cur}, \mathcal{D}_u^{cur}$, together with their history sequences $\mathcal{L}_u^{hist}, \mathcal{C}_u^{hist}, \mathcal{R}_u^{hist}, \mathcal{D}_u^{hist}$, and social relationship matrix $\mathbf{S}$}.

\begin{table}[t]
\centering
\addtolength{\tabcolsep}{-2pt}
\caption{Notations in MRP-LLM and their description} \label{tab:notation}
\begin{tabular}{l|l}
\specialrule{.15em}{.05em}{.05em}
\textbf{Notation} & \textbf{Description} \\
\specialrule{.1em}{.05em}{.05em}
$\mathcal{U}$ & user set \\
$\mathcal{L}$ & POI set\\
$\mathcal{C}$ & category set \\
$\mathcal{R}$ & region set \\
$t$ & timestamp of the check-in \\
$d$ & distance from the previous check-in \\
$\mathcal{L}_u$ & $u$'s POI check-in sequence \\
%
%
$\mathcal{C}_u$ & $u$'s category check-in sequence \\
%
%
$\mathcal{R}_u$ & $u$'s region check-in sequence \\
%
%
$\mathcal{D}_u$ & $u$'s distance check-in sequence \\
%
%
$\mathcal{X}^{cur}_u$ & $u$'s current sequence of any type $(\mathcal{L},\mathcal{C},\mathcal{R},\mathcal{D})$\\
$\mathcal{X}^{hist}_u$ & $u$'s history sequence of any type $(\mathcal{L},\mathcal{C},\mathcal{R},\mathcal{D})$\\
\textcolor{black}{$\mathbf{S}$} & \textcolor{black}{social relationship matrix of all users}\\
\specialrule{.15em}{.05em}{.05em}
\end{tabular}
\end{table}

\subsection{Framework Overview}
Figure~\ref{fig:1} presents an overview of MRP-LLM. Users' raw data are strictly kept on their local devices. Various types of user data are transmitted to different modules on the RS server with privacy protection in place. Specifically,
(1) Before recommendation, the \textit{Multitask Reflective Preference Extraction Module} distills each user’s fine-grained preferences from their $\mathcal{C}_u$, $\mathcal{R}_u$, and $\mathcal{D}_u$ via multitask preference probing and self-reflection. The preferences are stored in a preference knowledge base (KB) to enrich recommendation prompts.
(2) During recommendation, the \textit{Neighbor Preference Retrieval Module} first retrieves and summarizes relevant users' preferences to introduce the collaborative signal for the recommendation prompts. 
(3) The \textit{Multitask Next POI Recommendation Module} then generates next POI recommendations by aggregating both the user's and her neighbors' fine-grained preferences. 
(4) The \textit{Privacy Transmission Module} mitigates \textcolor{black}{users' check-in history and social relationship}
leakage during data uploading via different perturbation techniques.

\subsection{Multitask Reflective Preference Extraction Module}

This module aims to guide LLMs in understanding fine-grained user preferences to achieve more accurate recommendations. 
Research has identified key factors influencing users' POI choices: the function of the POI (category), spatial location (region and distance), and the temporal context of the check-in. the spatial location (region and distance), and the temporal context of the check-in (check-in order and time)~\cite{sun2023multi}. We reckon that temporal context may impact user preferences for category, region, and distance, which further breaks down preferences into five categories. 

Based on the temporal context, a user's choice of POI categories may depend on her (1) \textit{categorical transition preference}, indicating the order of category she prefers to visit e.g., restaurant $\rightarrow$ movie theatre, and her (2) \textit{categorical temporal preference}, indicating the time slot she prefers to visit a category, e.g., a restaurant at 6 pm. 
A user's choice of regions may also differ based on temporal contexts, resulting in the (3) \textit{regional transition preference}, indicating which regions the user tends to visit consecutively, e.g.,  $r_1\rightarrow r_2$, and the (4) \textit{regional temporal preference}, indicating which region the user tend to visit at a certain time, e.g., $r_1$ at 6 pm. 
Lastly, the time slot may affect a user's preferred traveling distance, i.e., the (5) \textit{distance temporal preference}. For instance, a user may commute a long distance at 9 am for work, and travel within 1km at 12 pm for lunch.

As direct prompting could not effectively identify such fine-grained user preferences~\cite{feng2024move}, we customize the prompts for next POI recommendation to probe user preferences step by step and refine iteratively. 

\subsubsection{Multitask Preference Probing}
We first use CoT to explicitly guide LLMs in capturing five types of user preferences across three subtasks.
%
Specifically, after providing the general instruction in Prompt 1, we present the LLM with $\mathcal{C}_u^{cur}$ and ask it to extract the user's categorical transition and temporal preferences in Prompt 2. Using the same approach, we then probe the user's regional transition and temporal preferences with $\mathcal{R}_u^{cur}$ and her distance temporal preference based on $\mathcal{D}_u^{cur}$, step by step.

\begin{tcolorbox}[title={Prompt 1: Task Instruction},
    boxsep=2pt,          
    left=2pt,            
    right=2pt,          
    top=2pt,             
    bottom=2pt]
    {\emph{Your task is to recommend a user's next point-of-interest (POI) from the candidate POIs \{\blue{$\mathcal{L}$}\} by analyzing the users' preferences on category, region, and distance.}}
\end{tcolorbox}\label{p1}

\begin{tcolorbox}[title={Prompt 2: Category Preference Probing},
    boxsep=2pt,          
    left=2pt,           
    right=2pt,          
    top=2pt,            
    bottom=2pt]
    {\emph{Given the user's Category sequence: \{\blue{$\mathcal{C}_u$}\}, what is the user's categorical transition preference? Considering: what are the `category pairs' the user usually visits consecutively? (format:\{category-category,...\})\\
    What is the user's categorical temporal preference? Considering: what are the `categories' the user visits at a certain time (day/ hour)? (format:\{time: [categories]\})}}
\end{tcolorbox}\label{p2}

\subsubsection{Dynamic Preference Self-Reflection.}

Multitask preference probing could extract more fine-grained user preferences. However, it still faces two limitations: (i) a single prompting may not capture the recent evolution of user preferences; and (ii) the context limit makes it infeasible to input long history sequences into LLMs for preference extraction.
As such, we introduce a dynamic preference self-reflection mechanism to address these challenges.

{First, to capture the recent evolution of user preferences, we sample the most recent $m$ segments, each of length $n$, from the current sequence as examples. Then the self-reflection mechanism~\cite{madaan2024self} is adopted to learn the recent preference changes from these segments.
For instance, for category preference reflection, we sample $m$ most recent segments $\mathcal{C}_u^{cur'}$ from $\mathcal{C}_u^{cur}$. 
For each $\mathcal{C}_u^{cur'}$, we send the segment, excluding the last record, to the LLM and ask it to predict the final category, as demonstrated in Prompt 3.
Prompt 4 then provides the LLM with the ground truth category and guides it to rectify its previous conclusions on the user's categorical transition and categorical temporal preferences.
A similar process will be conducted to update the user's region and distance preference reflection after the category preference reflection.
}

{Second, to integrate more information from the lengthy history sequence, we also sample $m$ segments of length $n$ each from the history sequence as examples to conduct the self-reflection.}
The segments are selected based on their contextual relevance to the recommendation task. 
Specifically, since the next POI recommendation task is to predict $l_{t_{k+1}}$ with current sequence $\{l_{t_1}, ... l_{t_{k-1}}, l_{t_{k}}\}$, any segment in the history sequence whose second last check-in is $l_{t_{k}}$ may have the highest context relevance to the recommendation task, followed by segments whose second last check-in is $l_{t_{k-1}}$, and so on.
We feed these $m$ segments to LLMs and perform preference probing and self-reflection with Prompts 2-4.

Ultimately, the five types of user preferences will be updated and stored in the fine-grained preference knowledge base (KB) for retrieval during the recommendation process.

\begin{tcolorbox}[title={Prompt 3: Category Preference Prediction},
    boxsep=2pt,        
    left=2pt,             
    right=2pt,            
    top=2pt,              
    bottom=2pt]
    {\emph{The user has visited categories \{\blue{$\mathcal{C}_u^{cur'}$}\}. Now is \{\blue{day}\} at \{\blue{hour}\}, based on the user's categorical transition preference and categorical temporal preference, predict users' next most likely visiting `category'. (format:category)}}
\end{tcolorbox}

\begin{tcolorbox}[title={Prompt 4: Category Preference Reflection},
    boxsep=2pt,          
    left=2pt,            
    right=2pt,           
    top=2pt,             
    bottom=2pt] 
    {\emph{The user actually visited category \{\blue{$c_{t_{k}}$}\}.\\
    Based on the actual visited category, what is the new insight you can get for the user's categorical transition preference? Generate the updated categorical transition preference. (format:\{category-category,...\})\\
    What is the new insight for the user's categorical temporal preference? Generate the updated categorical transition preference. (format:\{time: [categories]\})}}
\end{tcolorbox}

\subsection{Neighbor Preference Retrieval Module}
Solely relying on the user's own data fails to leverage global patterns. Thus, the Neighbor Preference Retrieval Module aims to leverage neighbors' preferences to introduce collaborative signals and further improve recommendation accuracy. 
We identify three types of neighbors. (1) \textit{Geographical neighbors} means users having similar preferences on geographical regions, symbolized by the Kullback-Leibler (KL) divergence~\cite{kullback1951information} of the regional check-in distribution, $RP(u)=\{\mathbb{P}(r_1), ..., \mathbb{P}(r_{\lvert\mathcal{R}\rvert})\}$,
\begin{equation}
    D_{\text{R}}(u_i, u_j) = \text{KL}(RP(u_i)\parallel RP(u_j));
\end{equation}
(2) \textit{Semantic neighbors} are users with similar category preferences, quantified by the KL divergence of check-in category distribution, $CP(u)=\{\mathbb{P}(c_1), \allowbreak ..., \allowbreak \mathbb{P}(c_{\lvert\mathcal{C}\rvert})\}$, 
\begin{equation}
    D_{\text{C}}(u_i, u_j) = \text{KL}(CP(u_i)\parallel CP(u_j));
\end{equation}
and (3) \textit{Social neighbors} are the social network neighbors, i.e., $D_{\text{S}}(u_i, u_j)=1$ if $u_i$ and $u_j$ are friends and vice versa.

Based on $D_{\text{R}}$, $D_{\text{C}}$, and $D_{\text{S}}$, it selects the closest neighbor of each type, retrieves their fine-grained preferences from the fine-grained preference KB, and leverages LLMs to summarize each of the five types of preferences with Prompt 5 (taking categorical transition preference as an example).

\begin{tcolorbox}[title={Prompt 5: Preference Summarization},
    boxsep=2pt,          
    left=2pt,            
    right=2pt,           
    top=2pt,             
    bottom=2pt]
    {\emph{The users' geographical neighbors' categorical transition preferences are \{\blue{preferences}\}; the users' semantic neighbors' categorical transition preferences are \{\blue{preferences}\}; the users' social neighbors' categorical transition preferences are \{\blue{preferences}\}. Summarize the neighbors' categorical transition preferences by considering their commonalities. \\
    (format: \{category-category,...\})} }
\end{tcolorbox}

\subsection{Multitask Next POI Recommendation Module}
This module aims to provide next POI recommendations by taking into account users' preferences on category, region, and distance.
Specifically, we first predict the next preferable category, region, and distance in three separate tasks with the guide of the user's and the neighbors' preferences, exemplified by Prompt 6. 
The predictions on subtasks provide hints on different aspects of user preferences. Finally, the LLM is asked to consider the preference hints and select the most suitable POI candidates using Prompt 7. Additionally, the relative importance of each aspect and an explanation of the recommendation are required to enhance interpretability.

\begin{tcolorbox}[title={Prompt 6: Next Category Prediction},
    boxsep=2pt,          
    left=2pt,            
    right=2pt,           
    top=2pt,             
    bottom=2pt]
    {\emph{Now is \{\blue{day}\} at \{\blue{hour}\}, based on the users' current category sequence \{\blue{$\mathcal{C}_u^{cur}$}\}, his own categorical transition preference and categorical temporal preference, and his neighbors' categorical transition preference and categorical temporal preference, predict the user's next most likely visiting `category'. (format: category)}}
\end{tcolorbox}

\begin{tcolorbox}[title={Prompt 7: Next POI Recommendation},
    boxsep=2pt,           
    left=2pt,            
    right=2pt,           
    top=2pt,             
    bottom=2pt]
    {\emph{Given users' current check-in sequence \{\blue{$\mathcal{L}_u^{cur}$}\}, 
    recommend 10 POIs from \{\blue{$\mathcal{L}$}\} considering his next likely visiting category, region, and distance. 
    State the reason for each recommendation and rank the importance of category, region, and distance preferences. \\
    (format: \{POI: reason; [importance ranking])\}}}
\end{tcolorbox}
\subsection{Privacy Transmission Module}
This module seeks to
preserve the privacy in four types of user data uploaded to the three modules of MRP-LLM by employing differential privacy techniques~\cite{dwork:2014algorithmic}. The detailed protection measures are described below.

(1) The Multitask Reflective Preference Extraction Module receives 
users' category, region, and distance sequences that could reveal users' preferences. 
Thus, we represent each record in a sequence by a one-hot vector and perturb it with optimized unary encoding (OUE) method~\cite{wang2017locally}. OUE algorithm perturbs the $i$-th bit of the original vector $\mathbf{x}$ into $\mathbf{x}'$ with privacy budget $\epsilon$ based on probability,
\begin{equation} \label{eq:OUE}
    \mathbb{P}(\mathbf{x}'[i]=1)= 
    \begin{cases}
    0.5, & \text{if } \mathbf{x}[i]=1\\
    1/(exp(\epsilon)+1), & \text{if } \mathbf{x}[i]=0 \text{.}
\end{cases}
\end{equation}

(2) The Neighbor Preference Retrieval Module requires category and region check-in distribution to identify semantic and geographical neighbors, $RP(u)$ and $CP(u)$, and users' \textcolor{black}{social relationship} for neighbor retrieval.
To protect the distribution, we inject noise sampled from the Laplace distribution. Specifically, we set new distribution $P'(u)$ based on original distribution $P(u)$ and privacy budget $\epsilon$,
\begin{equation} \label{eq:lap}
    P'(u) = P(u) + Lap(\Delta f/\epsilon),
\end{equation}
where the sensitivity $\Delta f$ in this case is $1$.
Furthermore, this module utilizes users' \textcolor{black}{social relationship} to seek their social neighbors. To protect user \textcolor{black}{social} links from exposing to unwanted personals or the recommender, we adopt the random flipping mechanism~\cite{gao2020dplcf} to perturb each \textcolor{black}{social} link of the \textcolor{black}{social relationship matrix $\mathbf{S}\in\mathbb{Z}^{|\mathcal{U}|\times|\mathcal{U}|}$.} 
Specifically, we first draw a random probability \textcolor{black}{$\phi \sim \text{Uniform}(0, 1)$} and then perturb the \textcolor{black}{social link $s_{ij}$} between $u_i$ and $u_j$ to \textcolor{black}{$s'_{ij}$} according to Equation~\ref{eq:flip}, given by,
%
\textcolor{black}{
\begin{equation} ~\label{eq:flip}
s'_{ij} = \begin{cases}
1, & \text{if } \phi \leq p \text{ and } s_{ij} = 1; \\
0, & \text{if } \phi > p \text{ and } s_{ij} = 1; \\
0, & \text{if } \phi > q \text{ and } s_{ij} = 0; \\
1, & \text{if } \phi \leq q \text{ and } s_{ij} = 0,
\end{cases}
\end{equation}
}
where $p/q\leq exp(\epsilon)$.

(3) \textcolor{black}{Multitask Next POI Recommendation Module} 
requires users' privacy-sensitive POI sequences for recommendation generation. To protect users' check-in records while retaining their geolocation proximity, we adapt the \textlangle $\varrho,h$\textrangle-privacy technique \textcolor{black}{~\cite{huo2021privacy}} to find a replacement POI near the check-in. 
For a user's check-in at $l$, we fuzzify it into a virtual circle whose center is $\delta$ away from $l$ and radius is $\varrho$. The value of $\varrho$ is determined by the check-in density $h$. To prevent some POIs in remote areas been unidentified due to a lack of neighbors, we further constrain the circle range \textcolor{black}{$\varrho$ to $[10km, 30km]$} 
to ensure the circle contains sufficient neighbors.
We then replace $l$ with another POI inside the virtual circle.
The algorithm of \textlangle$\varrho,h$\textrangle-privacy is shown in Algorithm~\ref{algo:1}.



    
    
    

    



\begin{algorithm}
    \caption{\textlangle$\varrho,h$\textrangle-Privacy for POI check-ins}
    \label{algo:1}
    
    \KwIn{$l$ with coordinate $(x_l, y_l)$ and category $c$, privacy parameter $\epsilon$}
    \KwOut{replacement POI $l'$}
    
    $h \leftarrow \text{random}(h_{min}, h_{max})$\;
    $\varrho \leftarrow \text{minimum radius containing at least } h \text{ POIs}$\;
    
    $\delta \leftarrow \text{random}(0,\varrho)$\;
    $\theta \leftarrow \text{random}(0, 2\pi)$\;
    
    $o \leftarrow \text{circle at } (x_l + \delta \cos \theta, y_l + \delta \sin \theta) \text{ with radius } \varrho$\;
    
    $c' \leftarrow \text{random\_flip}(\epsilon)$\;
    
    \eIf{$c' = 0$ \text{and} $o$ \text{contains POI of category} $c$}{
        $l' \leftarrow$ a random POI of category $c$ in $o$\;
    }{
        $l' \leftarrow$ a random POI in $o$\;
    }
    
    \Return $l'$\;

\end{algorithm}

\subsection{Summary}
The entire process for MRP-LLM to perform next POI recommendation task is summarized in Algorithm~\ref{algo:2}.
%
First, users upload their perturbed category, region, and distance sequences, together with their perturbed category, region distributions, and \textcolor{black}{social relationship} (lines 3-5). MRP-LLM then samples $m$ segments from uploaded sequences based on time and contextual relevance (line 15) and uses the segments to perform multiple rounds of multitask preference probing and dynamic preference self-reflection (lines 16-18). The resulting users' preferences are stored in the fine-grained preference KB (line 7). 
Second, when a user initiates a recommendation quest, MRP-LLM identifies her neighbors based on geographical, semantic, and social closeness, retrieves their preferences, and utilizes LLM to summarize their fine-grained preferences (lines 9-10).
Finally, the user's and her neighbors' preferences serve as important hints for LLM to perform the multitask next POI recommendation based on the user's perturbed current category, region, distance, and POI sequence (lines 11-13).

\begin{algorithm}
    \caption{Recommendation Process of MRP-LLM}
    \label{algo:2}
    
    \KwIn{$\mathcal{U}$, $\mathcal{L}_u$, $\mathcal{C}_u$, $\mathcal{R}_u$, $\mathcal{D}_u$}
    \KwOut{Next POI recommendation list $\mathbf{l}$}
    \SetKwFunction{FMain}{PreferenceExt} 
    \SetKwProg{Fn}{Function}{:}{end}  
    
    \tcp{Before recommendation request}
    Input general instruction (\textit{Prompt 1})\; 
    \For{$u$ in $\mathcal{U}$}{
        Collect $\mathcal{C}'_u$, $\mathcal{R}'_u$, $\mathcal{D}'_u$ perturbed as Equation~\ref{eq:OUE}\;
        Collect $RP'(u)$, $CP'(u)$ perturbed as Equation~\ref{eq:lap}\;
        Collect $\mathbf{S}'$ perturbed as Equation~\ref{eq:flip}\; 
        $Pref(u)\leftarrow$ \FMain{$\mathcal{C}'_u$, $\mathcal{R}'_u$, $\mathcal{D}'_u$}\;
        Store $Pref(u)$ into fine-grained KB\;
    }
    \tcp{{Upon each recommendation request}}
    \For{$u$ in $\mathcal{U}$}{
        Find $u$'s geographical, semantic, and social neighbors\;
        Summarize neighbor preferences (\textit{Prompt 5})\;
        Collect POI sequences perturbed by Algorithm~\ref{algo:1}\;
        $\mathbf{l} \leftarrow$ next POI recommendation (\textit{Prompts 6-7})\;
        \Return $\mathbf{l}$\;
    }
    \Fn{\FMain{$\mathcal{C}_u$, $\mathcal{R}_u$, $\mathcal{D}_u$}}{
        Select $m$ {segments} from $\mathcal{C}_u$, $\mathcal{R}_u$, $\mathcal{D}_u$\;
        \For{each example}{
            $Pref(u) \!\!\leftarrow\!\!$ preference extraction (\textit{Prompt 2})\; 
            $Pref(u) \!\!\leftarrow\!\!$ preference reflection (\textit{Prompts 3-4})\; 
        }
        \Return $Pref(u)$\;
    }

\end{algorithm}


\begin{table}[ht!]
\centering
\addtolength{\tabcolsep}{1pt}
\caption{Statistics of Data Utilized in Our Study.} 
\label{tab:1}
\begin{tabular}{l|l|l|l}
    \specialrule{.15em}{.05em}{.05em}
    City &SIN &NY & PHO\\
    \specialrule{.1em}{.05em}{.05em}
     \# Check-ins &355,337 &511,431 &47,980\\ 
     \# Users &8,648 &16,387 &2,946\\
     \# POIs &33,712 &56,252 &7,247 \\
     \# Categories &398 &420 &344\\
     Density &0.12\% &0.06\% &0.22\%\\
    \specialrule{.15em}{.05em}{.05em}
    \end{tabular}
\end{table}
\begin{table*}[ht!]
\centering
\caption{The search space of hyper-parameters and the optimal settings for all methods on the three real-world datasets.}\label{tab:hyper}
\begin{tabular}{c|l|l|l|l|l|l}
\specialrule{.15em}{.05em}{.05em}
\multicolumn{1}{c|}{Methods}  & Parameter     & SIN    & NY     & PHO     & Searching space         & Description                                       \\ \specialrule{.1em}{.1em}{.1em} 
\multirow{4}{*}{{BPRMF}}  
& -$d$ &  40  & 60  &  40   & {[}20, 100{]}, step size=20 & the size of embeddings \\
& -$lr$  &  1e-2  &  1e-2  &  1e-2  & \{1e-4, 1e-3, 1e-2\}   & learning rate  \\
& -$batch\_size$   &  16  &  64  &  64  & \{16, 32, 64, 128\}             & batch size\\ 
& -$\lambda$   &  1e-3  &   1e-2 &  1e-2  & \{1e-4, 1e-3, 1e-2\}             & L2 regularization coefficient \\
\specialrule{.05em}{.05em}{.05em}
\multirow{4}{*}{{STRNN}}  
& -$d$ &  80  & 60  &  100   &  {[}20, 100{]}, step size=20 & the size of embeddings\\
  & -$lr$  &  1e-3  &  1e-3  &  1e-3  & \{1e-4, 1e-3, 1e-2\}   & learning rate  \\
& -$batch\_size$   &  32  &  32  &  64  & \{16, 32, 64, 128\}             & batch size\\ 
& -$w$   &  1d  &  12h  &  2d  & \{6h, 12h, 1d, 2d, 3d\}            & width of time window \\
\specialrule{.05em}{.05em}{.05em}
\multirow{5}{*}{{STAN}}  & -$d$ &  60 &  70 &  50  & {[}20, 100{]}, step size=20 & the size of embeddings                   \\
  & -$lr$  &  1e-4  &  1e-4  &  1e-3  & \{1e-4, 1e-3, 1e-2\}   & learning rate  \\
& -$batch\_size$   &  64  &  64  &  128  & \{16, 32, 64, 128\}             & batch size\\ 
& -$negative\_size$   &  10  &  10 &  20  & {[}0, 50{]}, step size=10        & No. negative samples \\
& -$\eta$   &  0.2  &  0.3 &  0.1 & {[}0, 0.5{]}, step size=0.1        & drop out rate \\
\specialrule{.05em}{.05em}{.05em}
\multirow{6}{*}{{STHGCN}}  & -$d$ &  40  & 60  &  40   & {[}20, 100{]}, step size=20 & the size of embeddings                   \\
& -$lr$  &  1e-4  &  1e-4  &  1e-4  & \{1e-4, 1e-3, 1e-2\}   & learning rate  \\
& -$batch\_size$   &  16  &  32  &  32  & \{16, 32, 64, 128\}             & batch size\\ 
& -$\eta_{inter}$  &  0.005  &  0.005  &   0.01 & \{0.005, 0.01, 0.1, 0.5\}             & inter-user collaborative threshold\\ 
& -$\eta_{intra}$   &  0.0  &  0.01  &  0.01  & \{0.0, 0.01, 0.05, 0.2\}           & intra-user collaborative threshold\\
& -$sample\_size$   &  100  &  100  &  200  & \{100, 200, 300, 400\}             & one-hop sampling size\\ 
\specialrule{.05em}{.05em}{.05em}
\multirow{2}{*}{{LLMMob}}  & -$M$ &  40  & 40  &  40   & {[}10,50{]}, step size=10 & length of historical stays                   \\
& -$N$  &  6  &  5  &   5 & \{3,4,5,6,7\}   & length of context stays  \\
\specialrule{.05em}{.05em}{.05em}
\multirow{2}{*}{{LLMMove}}  & -$M$ &  30  & 40  &  40   & {[}10,50{]}, step size=10 & length of historical stays                   \\
& -$N$  &  5  &  3  &   5 & \{3,4,5,6,7\}   & length of context stays  \\
\specialrule{.05em}{.05em}{.05em}
\multirow{4}{*}{{MRP-LLM}}  
& -$m$ &  1  &  1 &   2  & \{1,2,3\} & number of examples \\
& -$n$  &  5  &  3  &  5  & {[}3,7{]}, step size=1   & maximum length of examples \\
& -$\rho$   &  1  & 1   &  1  & {[}0.25, 1{]}, step size=0.25  & neighbor participation rate \\
& -$\epsilon$   &  0.1  &  0.1  &   0.1 & {[}0.1, 0.9{]}, step size=0.2     & privacy budget\\
\specialrule{.15em}{.05em}{.05em}
\end{tabular}
\end{table*}

\section{Experiments and Analysis}\label{sec:experiments}

\begin{table*}[ht!]
\centering
\renewcommand{\arraystretch}{1.1}
\addtolength{\tabcolsep}{2pt}
\caption{Performance comparison of conventional (rows 1-6) and LLM-based baselines (rows 7-8) with our methods with/without privacy preservation (rows 9-10) \textcolor{black}{on SIN, NY, and PHO}. We repeat each evaluation 10 times and report the average result.
}
\label{table:result_SIN}
\begin{tabular}{c|c|l|c c c c} 
\specialrule{.15em}{.05em}{.05em}
    SIN &Row ID &Methods & ACC@1 & ACC@5 & ACC@10 & MRR \\
    \specialrule{.1em}{.05em}{.05em}
    {\multirow{6}{*}{{\textbf{Conventional Methods}}}} &\textit{1} & MostPop & 0.0550  & 0.2050 & 0.2600 & 0.0973 \\
    &\textit{2} & Dist & 0.1700  & 0.2700 & 0.3150 & 0.2452 \\
    &\textit{3} & BPRMF & 0.2100  & 0.3050 & 0.3850 & 0.2477 \\
    &\textit{4} & STRNN & 0.2050  & 0.3500 & 0.4100 & 0. 2764  \\
    &\textit{5} & STAN  & 0.6000  & 0.6800$^\dagger$ & 0.7400 & 0.6373$^\dagger$ \\
    &\textit{6} & STHGCN & 0.6150$^\dagger$ & 0.6550 & 0.7600$^\dagger$ & 0.6313 \\
    \hline
    {\multirow{2}{*}{{\textbf{LLM-based Methods}}}} &\textit{7} & LLMMob & 0.3850  & 0.5850 & 0.6550 & 0.4384 \\
    &\textit{8} & LLMMove & 0.5250$^\ddagger$  & 0.6100$^\ddagger$ & 0.6600$^\ddagger$ & 0.5474$^\ddagger$ \\
    \specialrule{.05em}{.05em}{.05em}
    {\multirow{2}{*}{
    {\textbf{Our Proposed Methods}}}}
    &\textit{9} & MR-LLM & 0.6000$^{*}$  & 0.6500$^{*}$ & 0.7450$^{*}$ & 0.6138$^{*}$ \\
    &\textit{10} & MRP-LLM & 0.5600$^{**}$  & 0.6100$^{**}$ & 0.6500$^{**}$ & 0.5762$^{**}$ \\
    \specialrule{.05em}{.05em}{.05em}
    {\multirow{4}{*}{{\textbf{Improvements}}}} 
    &\textit{11} & $*$ vs. $\dagger$ & -2.4\%  & -4.4\% & -2.0\% & -3.7\% \\
    &\textit{12} & $*$ vs. $\ddagger$ & 14.2\%  & 6.6\% & 12.9\% & 12.1\% \\
    &\textit{13} & $**$ vs. $\dagger$ & -8.9\%  & -10.3\% & -14.5\% & -9.6\% \\
    &\textit{14} & $**$ vs. $\ddagger$ & 6.7\%  & 0\% & -1.5\% & 5.3\% \\
    \specialrule{.15em}{.05em}{.05em} 
\specialrule{.15em}{.05em}{.05em}
    NY& Row ID & Methods & ACC@1 & ACC@5 & ACC@10 & MRR \\
    \specialrule{.1em}{.05em}{.05em}
    {\multirow{6}{*}{{\textbf{Conventional Methods}}}} &\textit{1} & MostPop & 0.0300 & 0.1450 & 0.1950 & 0.0729 \\
    &\textit{2} & Dist & 0.0700 & 0.1650 & 0.2300 & 0.1052 \\
    &\textit{3} & BPRMF & 0.1600 & 0.2550 & 0.3550 & 0.2077 \\
    &\textit{4} & STRNN & 0.1750 & 0.3000 & 0.3900 & 0.2317 \\
    &\textit{5} & STAN & 0.4750 & 0.5800 & 0.6550$^\dagger$ & 0.5135 \\
    &\textit{6} & STHGCN & 0.5000$^\dagger$  & 0.5850$^\dagger$ & 0.6500 & 0.5346$^\dagger$ \\
    \hline
    {\multirow{2}{*}{
    {\textbf{Conventional Methods}}}} &\textit{7} & LLMMob & 0.3050  & 0.4600 & 0.4950 & 0.3544 \\
    &\textit{8} & LLMMove & 0.4300$^\ddagger$ & 0.5400$^\ddagger$ & 0.5850$^\ddagger$ & 0.4856$^\ddagger$ \\
    \specialrule{.05em}{.05em}{.05em}
    {\multirow{2}{*}{
    {\textbf{Our Proposed Methods}}}} 
    &\textit{9} & MR-LLM & 0.4650$^{*}$ & 0.5650$^{*}$ & 0.6050$^{*}$ & 0.5036$^{*}$\\
    &\textit{10} & MRP-LLM & 0.3950$^{**}$ & 0.5050$^{**}$ & 0.5550$^{**}$ & 0.4552$^{**}$ \\  
    \specialrule{.05em}{.05em}{.05em}
    {\multirow{4}{*}{{\textbf{Improvements}}}} 
    &\textit{11} & $*$ vs. $\dagger$ & -7.0\% & -3.4\% & -7.6\% & -5.8\% \\
    &\textit{12} & $*$ vs. $\ddagger$ & 8.1\% & 4.6\% & 3.4\% & 3.7\% \\
    &\textit{13} & $**$ vs. $\dagger$ & -21\% & -13.7\% & -15.3\% & -14.9\% \\
    &\textit{14} & $**$ vs. $\ddagger$ & -8.1\% & -6.5\% & -5.1\% & -6.3\% \\
    \specialrule{.15em}{.05em}{.05em}
\specialrule{.15em}{.05em}{.05em}
    PHO & Row ID & Methods & ACC@1 & ACC@5 & ACC@10 & MRR \\
    \specialrule{.1em}{.05em}{.05em}
    {\multirow{6}{*}{{\textbf{Conventional Methods}}}} &\textit{1} & MostPop & 0.1400  & 0.2550 & 0.3000 & 0.1960\\
    &\textit{2} & Dist & 0.2050  & 0.3050 & 0.3600 & 0.2766\\
    &\textit{3} & BPRMF & 0.2400  & 0.3500 & 0.4250 & 0.3216\\
    &\textit{4} & STRNN & 0.2850  & 0.4000 & 0.4850 & 0.3742\\
    &\textit{5} & STAN & 0.6200  & 0.6950$^\dagger$ & 0.7750 & 0.6587\\
    &\textit{6} & STHGCN & 0.6300$^\dagger$  & 0.6800 & 0.7950$^\dagger$ & 0.6662$^\dagger$\\
    \hline
    {\multirow{2}{*}{
    {\textbf{Conventional Methods}}}} &\textit{7} & LLMMob & 0.4000  & 0.5800 & 0.6750 & 0.4610\\
    &\textit{8} & LLMMove  & 0.5350$^\ddagger$  & 0.6450$^\ddagger$ & 0.6900$^\ddagger$ & 0.5829$^\ddagger$\\
    \specialrule{.05em}{.05em}{.05em}
    {\multirow{2}{*}{
    {\textbf{Our Proposed Methods}}}} 
    &\textit{9} & MR-LLM & 0.6250$^{*}$  & 0.6550$^{*}$ & 0.7400$^{*}$ & 0.6138$^{*}$\\
    &\textit{10} & MRP-LLM & 0.5800$^{**}$  & 0.6100$^{**}$ & 0.6900$^{**}$ & 0.6033$^{**}$\\
    \specialrule{.05em}{.05em}{.05em}
    {\multirow{4}{*}{{\textbf{Improvements}}}} 
    &\textit{11} & $*$ vs. $\dagger$ & -0.8\% & -5.8\%  & -6.9\% & -7.9\%\\
    &\textit{12} & $*$ vs. $\ddagger$ & 16.8\%  & 1.6\% & 7.2\% & 5.3\%\\
    &\textit{13} & $**$ vs. $\dagger$ & -7.9\% & -12.2\%  & -13.2\% & -9.4\% \\
    &\textit{14} & $**$ vs. $\ddagger$ & 8.4\%  & -5.4\% & 0\% & 3.5\%\\
    \specialrule{.15em}{.05em}{.05em} 
    \end{tabular} 
\end{table*}

We conducted experiments on three real-world datasets to answer the following four research questions.  
\begin{itemize}
    \item (\textbf{RQ1}): How does the recommendation accuracy of MRP-LLM compare to existing conventional and LLM-based next POI RSs?
    \item (\textbf{RQ2}): How does each component affect the performance of our proposed MRP-LLM? 
    \item (\textbf{RQ3}): How do hyper-parameters affect the performance of our proposed MRP-LLM?
    \item (\textbf{RQ4}): How reasonable and interpretable is the output of our proposed MRP-LLM?
\end{itemize}

\subsection{Datasets}
We utilized the Foursquare dataset~\cite{yang2016participatory} in three cities, namely Singapore (SIN), New York (NY), and Phoenix (PHO) constituting user check-in history, POI geographical location, and user friendships. Details are shown in Table~\ref{tab:1}.
Following existing works~\cite{zhang:2022next}, we preprocess the dataset first by applying 5-core filtering. For each user, we split the check-ins by day and filter out users with less than 3 sequences. The sequences are then chronologically split into training, validation, and test sets with a ratio of 8:1:1. 
It is noteworthy that since in MRP-LLM, we are performing ICL 
without the need for model training, the training set is used as history sequences. For validation and test sets, we leave the last check-in as ground truth and use the previous check-ins as the current sequence.

\begin{figure*}[t]
\centering
\subfigure{
\begin{tikzpicture}[xscale=0.7,yscale=0.7]
\begin{axis}[
    title={(a) SIN},
    title style={at={(0.5,-0.3)},anchor=north,font=\large},
    xtick align=inside,
    width=1.4\textwidth,
    height=0.25\textwidth,
    ybar,
    bar width=4.5pt,
    symbolic x coords={MRP-LLM, pos1, -MP, -MP-C, -MP-R, -MP-D, pos2, -SR, -SR-R, -SR-H, pos3, -NR, -NR-G, -NR-C, -NR-S, pos4, -PT, -PT-S, -PT-D, -PT-P},
    tick label style={font=\normalsize},
    yticklabel style={/pgf/number format/.cd,fixed,precision=3},
    ymin=0, ymax=1,
    enlarge x limits=0.05,
    xtick={MRP-LLM, -MP, -MP-C, -MP-R, -MP-D, -SR, -SR-R, -SR-H, -NR, -NR-G, -NR-C, -NR-S, -PT, -PT-S, -PT-D, -PT-P},
    xlabel style={font=\normalsize},
    ylabel style={font=\normalsize},
    ylabel near ticks,
    scaled ticks=false,
    ymajorgrids=true,
    grid style=dashed,
    legend style={at={(0.5,0.95)}, anchor=north, legend columns=4, draw=none, column sep=0.3in, font=\normalsize}, 
    legend entries={ACC@1, ACC@5, ACC@10, MRR}, 
    ]
    \addplot [color=orange,fill=orange!45, opacity=0.6 ] coordinates {
        (MRP-LLM, 0.56) 
        (pos1, 0) 
        (-MP, 0.42)
        (-MP-C, 0.45)
        (-MP-R, 0.47)
        (-MP-D, 0.48)
        (pos2, 0)
        (-SR, 0.515)
        (-SR-R, 0.535)
        (-SR-H, 0.55)
        (pos3, 0)
        (-NR, 0.49)
        (-NR-G, 0.51)
        (-NR-C, 0.515)
        (-NR-S, 0.51)
        (pos4,0)
        (-PT, 0.6)
        (-PT-S, 0.57)
        (-PT-D, 0.575)
        (-PT-P, 0.585)
        
    };
    \addplot [color=gray,fill=gray!45, opacity=0.6 ] coordinates {
        (MRP-LLM, 0.61) 
        (pos1, 0) 
        (-MP, 0.47)
        (-MP-C, 0.49)
        (-MP-R, 0.55)
        (-MP-D, 0.6)
        (pos2, 0)
        (-SR, 0.57)
        (-SR-R, 0.6)
        (-SR-H, 0.595)
        (pos3, 0)
        (-NR, 0.545)
        (-NR-G, 0.56)
        (-NR-C, 0.58)
        (-NR-S, 0.58)
        (pos4,0)
        (-PT, 0.65)
        (-PT-S, 0.625)
        (-PT-D, 0.615)
        (-PT-P, 0.64)
        
    };
    \addplot [color=blue,fill=blue!45, opacity=0.6 ] coordinates {
        (MRP-LLM, 0.6500) 
        (pos1, 0) 
        (-MP, 0.5300)
        (-MP-C, 0.5500)
        (-MP-R, 0.5950)
        (-MP-D, 0.6350)
        (pos2, 0)
        (-SR, 0.6150)
        (-SR-R, 0.6300)
        (-SR-H, 0.6250)
        (pos3, 0)
        (-NR, 0.5950)
        (-NR-G, 0.6000)
        (-NR-C, 0.6300)
        (-NR-S, 0.6400)
        (pos4,0)
        (-PT, 0.7450)
        (-PT-S, 0.6750)
        (-PT-D, 0.6550)
        (-PT-P, 0.7200)
        
    };
    \addplot [color=purple,fill=purple!45, opacity=0.6 ] coordinates {
        (MRP-LLM, 0.5762) 
        (pos1, 0) 
        (-MP, 0.4533)
        (-MP-C, 0.4792)
        (-MP-R, 0.5139)
        (-MP-D, 0.5572)
        (pos2, 0)
        (-SR, 0.5388)
        (-SR-R, 0.5598)
        (-SR-H, 0.5789)
        (pos3, 0)
        (-NR, 0.5103)
        (-NR-G, 0.5288)
        (-NR-C, 0.5367)
        (-NR-S, 0.531)
        (pos4,0)
        (-PT, 0.6138)
        (-PT-S, 0.5903)
        (-PT-D, 0.5847)
        (-PT-P, 0.5968)        
    };
\end{axis}
\end{tikzpicture}}
\vspace{-0.1in}

\subfigure{
\begin{tikzpicture}[xscale=0.7,yscale=0.7]
\begin{axis}[
    title={(b) NY},
    title style={at={(0.5,-0.3)},anchor=north,font=\large},
    xtick align=inside,
    width=1.4\textwidth,
    height=0.25\textwidth,
    ybar,
    bar width=4.5pt,
    symbolic x coords={MRP-LLM, pos1, -MP, -MP-C, -MP-R, -MP-D, pos2, -SR, -SR-R, -SR-H, pos3, -NR, -NR-G, -NR-C, -NR-S, pos4, -PT, -PT-S, -PT-D, -PT-P},
    tick label style={font=\normalsize},
    yticklabel style={/pgf/number format/.cd,fixed,precision=3},
    ymin=0, ymax=1,
    enlarge x limits=0.05,
    xtick={MRP-LLM, -MP, -MP-C, -MP-R, -MP-D, -SR, -SR-R, -SR-H, -NR, -NR-G, -NR-C, -NR-S, -PT, -PT-S, -PT-D, -PT-P},
    xlabel style={font=\normalsize},
    ylabel style={font=\normalsize},
    ylabel near ticks,
    scaled ticks=false,
    ymajorgrids=true,
    grid style=dashed,
    legend style={at={(0.5,0.95)}, anchor=north, legend columns=4, draw=none, column sep=0.3in, font=\normalsize}, 
    legend entries={ACC@1, ACC@5, ACC@10, MRR}, 
    ]
    \addplot [color=orange,fill=orange!45, opacity=0.6 ] coordinates {
        (MRP-LLM, 0.395) 
        (pos1, 0) 
        (-MP, .0355)
        (-MP-C, 0.375)
        (-MP-R, 0.38)
        (-MP-D, 0.38)
        (pos2, 0)
        (-SR, 0.395)
        (-SR-R, 0.45)
        (-SR-H, 0.425)
        (pos3, 0)
        (-NR, 0.36)
        (-NR-G, 0.38)
        (-NR-C, 0.405)
        (-NR-S, 0.4)
        (pos4,0)
        (-PT, 0.465)
        (-PT-S, 0.41)
        (-PT-D, 0.425)
        (-PT-P, 0.455)
        
    };
    \addplot [color=gray,fill=gray!45, opacity=0.6 ] coordinates {
        (MRP-LLM, 0.505) 
        (pos1, 0) 
        (-MP, 0.45)
        (-MP-C, 0.47)
        (-MP-R, 0.485)
        (-MP-D, 0.47)
        (pos2, 0)
        (-SR, 0.485)
        (-SR-R, 0.495)
        (-SR-H, 0.5)
        (pos3, 0)
        (-NR, 0.465)
        (-NR-G, 0.485)
        (-NR-C, 0.495)
        (-NR-S, 0.49)
        (pos4,0)
        (-PT, 0.565)
        (-PT-S, 0.53)
        (-PT-D, 0.525)
        (-PT-P, 0.55)
        
    };
    \addplot [color=blue,fill=blue!45, opacity=0.6 ] coordinates {
        (MRP-LLM, 0.555) 
        (pos1, 0) 
        (-MP, 0.505)
        (-MP-C, 0.535)
        (-MP-R, 0.55)
        (-MP-D, 0.54)
        (pos2, 0)
        (-SR, 0.51)
        (-SR-R, 0.535)
        (-SR-H, 0.54)
        (pos3, 0)
        (-NR, 0.53)
        (-NR-G, 0.54)
        (-NR-C, 0.545)
        (-NR-S, 0.54)
        (pos4,0)
        (-PT, 0.605)
        (-PT-S, 0.57)
        (-PT-D, 0.585)
        (-PT-P, 0.6)
        
    };
    \addplot [color=purple,fill=purple!45, opacity=0.6 ] coordinates {
        (MRP-LLM, 0.4552) 
        (pos1, 0) 
        (-MP, 0.4101)
        (-MP-C, 0.423)
        (-MP-R, 0.4279)
        (-MP-D, 0.4322)
        (pos2, 0)
        (-SR, 0.4312)
        (-SR-R, 0.4762)
        (-SR-H, 0.4665)
        (pos3, 0)
        (-NR, 0.4067)
        (-NR-G, 0.4301)
        (-NR-C, 0.4389)
        (-NR-S, 0.4407)
        (pos4,0)
        (-PT, 0.5036)
        (-PT-S, 0.4587)
        (-PT-D, 0.4504)
        (-PT-P, 0.4907)
        
    };
\end{axis}
\end{tikzpicture}}
\vspace{-0.1in}

\subfigure{
\begin{tikzpicture}[xscale=0.7,yscale=0.7]
\begin{axis}[
    title={(c) PHO},
    title style={at={(0.5,-0.3)},anchor=north,font=\large},
    xtick align=inside,
    width=1.4\textwidth,
    height=0.25\textwidth,
    ybar,
    bar width=4.5pt,
    symbolic x coords={MRP-LLM, pos1, -MP, -MP-C, -MP-R, -MP-D, pos2, -SR, -SR-R, -SR-H, pos3, -NR, -NR-G, -NR-C, -NR-S, pos4, -PT, -PT-S, -PT-D, -PT-P},
    tick label style={font=\normalsize},
    yticklabel style={/pgf/number format/.cd,fixed,precision=3},
    ymin=0, ymax=1,
    enlarge x limits=0.05,
    xtick={MRP-LLM, -MP, -MP-C, -MP-R, -MP-D, -SR, -SR-R, -SR-H, -NR, -NR-G, -NR-C, -NR-S, -PT, -PT-S, -PT-D, -PT-P},
    xlabel style={font=\normalsize},
    ylabel style={font=\normalsize},
    ylabel near ticks,
    scaled ticks=false,
    ymajorgrids=true,
    grid style=dashed,
    legend style={at={(0.5,0.95)}, anchor=north, legend columns=4, draw=none, column sep=0.3in, font=\normalsize}, 
    legend entries={ACC@1, ACC@5, ACC@10, MRR}, 
    ]
    \addplot [color=orange,fill=orange!45, opacity=0.6 ] coordinates {
        (MRP-LLM, 0.58) 
        (pos1, 0) 
        (-MP, 0.505)
        (-MP-C, 0.535)
        (-MP-R, 0.54)
        (-MP-D, 0.56)
        (pos2, 0)
        (-SR, 0.535)
        (-SR-R, 0.55)
        (-SR-H, 0.545)
        (pos3, 0)
        (-NR, 0.55)
        (-NR-G, 0.555)
        (-NR-C, 0.565)
        (-NR-S, 0.57)
        (pos4,0)
        (-PT, 0.625)
        (-PT-S, 0.59)
        (-PT-D, 0.59)
        (-PT-P, 0.61)
        
    };
    \addplot [color=gray,fill=gray!45, opacity=0.6 ] coordinates {
        (MRP-LLM, 0.61) 
        (pos1, 0) 
        (-MP, 0.54)
        (-MP-C, 0.56)
        (-MP-R, 0.59)
        (-MP-D, 0.575)
        (pos2, 0)
        (-SR, 0.57)
        (-SR-R, 0.58)
        (-SR-H, 0.595)
        (pos3, 0)
        (-NR, 0.575)
        (-NR-G, 0.57)
        (-NR-C, 0.605)
        (-NR-S, 0.6)
        (pos4,0)
        (-PT, 0.655)
        (-PT-S, 0.62)
        (-PT-D, 0.615)
        (-PT-P, 0.64)
        
    };
    \addplot [color=blue,fill=blue!45, opacity=0.6 ] coordinates {
        (MRP-LLM, 0.6900) 
        (pos1, 0) 
        (-MP, 0.57)
        (-MP-C, 0.5850)
        (-MP-R, 0.605)
        (-MP-D, 0.59)
        (pos2, 0)
        (-SR, 0.605)
        (-SR-R, 0.6550)
        (-SR-H, 0.63)
        (pos3, 0)
        (-NR, 0.605)
        (-NR-G, 0.62)
        (-NR-C, 0.645)
        (-NR-S, 0.64)
        (pos4,0)
        (-PT, 0.7400)
        (-PT-S, 0.715)
        (-PT-D, 0.71)
        (-PT-P, 0.73)
        
    };
    \addplot [color=purple,fill=purple!45, opacity=0.6 ] coordinates {
        (MRP-LLM, 0.6033) 
        (pos1, 0) 
        (-MP, 0.5239)
        (-MP-C, 0.5497)
        (-MP-R, 0.567)
        (-MP-D, 0.5703)
        (pos2, 0)
        (-SR, 0.5504)
        (-SR-R, 0.5726)
        (-SR-H, 0.5803)
        (pos3, 0)
        (-NR, 0.563)
        (-NR-G, 0.5684)
        (-NR-C, 0.5872)
        (-NR-S, 0.5897)
        (pos4,0)
        (-PT, 0.6138)
        (-PT-S, 0.6177)
        (-PT-D, 0.6132)
        (-PT-P, 0.6208)
        
    };
\end{axis}
\end{tikzpicture}}
\caption{Results of ablation study.}
\label{fig:3}
\end{figure*}
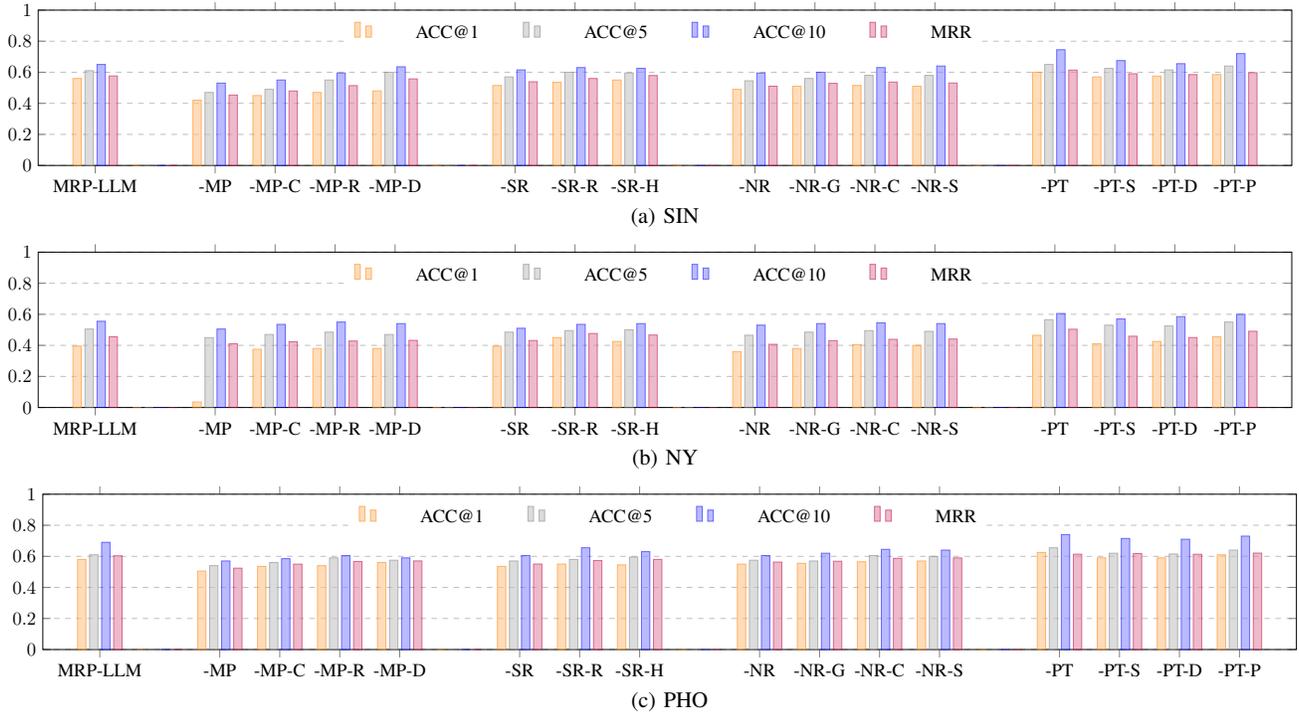

\subsection{Evaluation Metrics} 
In line with SOTA approaches~\cite{wang2023would,feng2024move}, we use Accuracy (ACC@K) and Mean Reciprocal Rank (MRR) to evaluate all methods. In general, higher metric values indicate better ranking performance. To ensure robust comparisons, we repeated each evaluation 10 times and reported the average result.

\subsection{Compared Baselines} We compare MRP-LLM with eight baselines for conventional next POI RSs (1-6) and LLM-based next POI RSs (7-8), 
\begin{enumerate}
    \item \textit{\textbf{MostPop}:} recommending the most popular POIs; 
    \item \textit{\textbf{Dist}:} selecting the nearest POIs; 
    \item \textit{\textbf{BPRMF}:}~\cite{rendle:2012bpr} performing matrix factorization on check-in history;
    \item \textit{\textbf{STRNN}:}~\cite{liu:2016predict} an RNN-based model incorporating check-in time interval and distance; 
    \item \textit{\textbf{STAN}:}~\cite{luo:2021stan} a bi-attention model that exploits the check-in spatiotemporal correlation;
    \item \textit{\textbf{STHGCN}:}~\cite{yan2023spatio} a hypergraph model leveraging POI, category, and spatiotemporal differences.
\end{enumerate}
%
%
%
%
 
%
\begin{enumerate}
\setcounter{enumi}{6}
    \item \textit{\textbf{LLMMob}:}~\cite{wang2023would} performing zero-shot recommendation via LLMs based on users' history and current sequences;
    \item \textit{\textbf{LLMMove}:}~\cite{feng2024move} performing zero-shot recommendation using LLMs based on users' check-in sequences, distance, and transition patterns. 
\end{enumerate}

\subsection{Implementation Details} We empirically find all the best hyper-parameters for all methods on the three datasets. 
For each evaluation, we randomly sample 100 POIs containing the ground truth as candidate sets. 
%
%
For LLM-based methods, we utilize \textit{gpt-3.5-turbo} as the LLM engine.
The detailed parameter searching space and best parameter settings for all methods are reported in Table~\ref{tab:hyper} to support reproducibility. 


\input{Plots/sensitivity}

\subsection{Model Performance Comparison (RQ1)}\label{sec:results}

Tables~\ref{table:result_SIN} presents the performance of all baselines and our methods across the three datasets, where `$\dagger$' marks the best results achieved by the conventional baselines; `$\ddagger$' highlights the best results achieved for LLM-based baselines. 
\textcolor{black}{Our method without privacy preservation (MR-LLM) is marked with `*', and the version with privacy preservation (MRP-LLM) is marked with `**'.
`Improvements' indicates the relative improvements between two methods, e.g., `* vs. $\ddagger$' in row 12 is the improvements comparing the results marked with `*' (achieved by MR-LLM) and those marked with `$\ddagger$' (obtained by the best LLM-based baseline).}

Several observations are noted.
\textbf{(1)} SOTA conventional POI recommendation methods, such as STAN and STHGCN trained with full-size data, achieve the highest accuracy among all approaches. However, our method without privacy-preserving (MR-LLM), despite not requiring the resource-intensive fine-tuning process, significantly reduces the performance drop, \textcolor{black}{limiting the average decrease to 4.5\% and 5.8\% in terms of ACC and MRR, respectively ({see row 11 `$*$ vs $\dagger$'})}.
%
\textbf{(2)} For LLM-based methods, LLMMove performs better than LLMob by introducing geographical information. Our MR-LLM greatly outperforms LLMMove with \textcolor{black}{an average lift of $8.4\%$ and $7.0\%$ w.r.t. ACC and MRR (see row 12 `$*$ vs $\ddagger$').} Even with privacy preservation, MRP-LLM could achieve similar performance compared to LLMMove, with \textcolor{black}{a slight drop of $1.3\%$ and a lift of $0.8\%$ w.r.t. ACC and MRR (see row 14  `$**$ vs $\ddagger$').}
\textbf{(3)} Overall, MR-LLM achieves outstanding performance with ICL compared with other LLM-based next POI RSs, and possesses comparable performance even with conventional next POI RSs trained with the full-size data. Meanwhile, MRP-LLM achieves the best balance between privacy and utility, achieving on-par performance with non-privacy-preserving LLM-based SOTAs.

\subsection{Components Effectiveness Analysis (RQ2)}
We examine the efficacy of different components of MRP-LLM by removing each of them.
\textbf{First}, we remove multitask preference probing (-MP). Within this component, we also remove each subtask to examine their impacts, i.e., categroy (-MP-C), region (-MP-R), and distance (-MP-D) probing, respectively.
\textbf{Second}, we remove the self-reflective component (-SR). Within the component, we also remove each type of segment used for reflection, i.e., segments drawn from recent sequences (-SR-R) and history sequences (-SR-H). 
\textbf{Third}, We remove the Neighbor Preference Retrieval Module (-NR). Within this component, we further remove each type of neighbor, i.e., geographical (-NR-G), semantic (-NR-C), and social neighbors (-NR-S).
\textbf{Last}, We remove the Privacy Transmission Module (-PT). Within this module, we also remove perturbation on each type of data: the category, region, and distance sequences (-PT-S), the category and region distributions (-PT-D), and the POI sequences (-PT-P).

Figure~\ref{fig:3} reveals several key observations based on the empirical results.
\textbf{(1)} MP improves model accuracy. Among them, the subtask on category has the largest impact on most occasions, followed by region and distance;
\textbf{(2)} SR enhances the performance, where temporal and contextual segments can lead to comparable improvements.
\textbf{(3)} NR can effectively enhance recommendation accuracy, where the geographical neighbors have the highest contribution, followed by semantic and social neighbors;
\textbf{(4)} In PT, the protection on POI sequences degrades accuracy the most, while perturbing distributions have the least impact on performance.

\subsection{Parameter Sensitivity Analysis (RQ3)}
Figures~\ref{fig:4m}-\ref{fig:4eps} depict the impact of essential hyper-parameters, including the segment number $m$ and length $n$, the preference extraction participation rate $\rho$, i.e., how many percentages of users upload their sequences for multitask preference extraction and then save their preferences to the KB for neighbor retrieval, and the privacy budget $\epsilon$.
Several findings are noted. 
\textbf{(1)} There is no obvious consistent trend on the impact of $m$ on accuracy across all datasets. Considering efficiency, we believe choosing one for each type of segment is optimal.
\textbf{(2)} The optimal length of each segment is $n=5$, which can provide sufficient context and avoid irrelevant noise.
\textbf{(3)} MRP-LLM performs the best when $\rho=100\%$. However, it could achieve comparable accuracy even when only $50\%$ users upload their preferences, indicating its robustness.
\textbf{(4)} A higher $\epsilon$ results in less noise adding to user data and increased accuracy, illustrating the privacy-utility trade-off.

\subsection{Recommendation Case Study (RQ4)} 
To directly examine the effectiveness and interpretability of MRP-LLM, we demonstrate a case study at each stage of a real recommendation process by randomly sampling a user on PHO dataset. Accordingly, it can be observed that 
\textbf{(1)} at stage 1, the user's category preferences can be accurately captured during preference probing;
\textbf{(2)} at stage 2, the user preference can be correctly updated by showing sequence segments as examples;
\textbf{(3)} at stage 3, neighbors' preferences can be summarized and even combined to form new relevant preferences, e.g., the `department store$\rightarrow$restaurants' transition;
and \textbf{(4)} at stage 4, the LLM accurately infers the next category using collaborative signals from neighbors, while effectively weighing different aspects of user preferences to provide well-reasoned recommendations. 

\tcbset{
    enhanced,
    colframe=ForestGreen, 
    colback=white, 
    coltitle=white, 
    fonttitle=\bfseries,
    title={},
    boxrule=0.5mm, 
    rounded corners,
    width=0.95\linewidth, 
    arc=1mm, 
    before=\vspace{0.5em}, 
    after=\vspace{0.7em}, 
    boxsep=0pt, 
    left skip=0pt, 
    right skip=5pt,
    top=3mm,                
    bottom=3mm,             
    left=5mm,               
    right=5mm,              
    breakable,
}

    \begin{tcolorbox}[title={Stage 1: Multitask Preference Probing},
        boxsep=2pt,           
        left=2pt,             
        right=2pt,            
        top=2pt,              
        bottom=2pt]
        {
        \textbf{Prompt:}
        \textlangle Prmopt 1\textrangle. Given the user’s Category sequence: \{\blue{(Restaurants, Sun, 1am), (Restaurants, Sun, 2am), (Bars, Sun, 3am), (Pet Services, Sun, 14pm), (Restaurants, Sun, 6pm), (Hotel, Sun, 9pm)}\}
        what is the user’s categorical transition preference? Considering: what are the ‘category pairs’ the user usually visits consecutively? (format:{category-category,...})
        
        \textbf{Output:}
        \emph{\textbf{\{{Restaurants-Bars, Bars-Pet Services, Restaurants -Hotel}\}}.}
    
        \textbf{Prompt:} What is the user's categorical temporal preference? Considering: what are the `categories' the user visits at a certain time (day/ hour)? (format:\{time: [categories]\})
        
        \textbf{Output:} 
        \emph{\textbf{{\{Early Morning: [Restaurants], Morning: [Hotel], Afternoon: [Pet Services], Evening: [Bars, Restaurants]\}}}}}
    \end{tcolorbox}

    \begin{tcolorbox}[title={Stage 2: Dynamic Preference Self-Reflection},
        boxsep=2pt,           
        left=2pt,             
        right=2pt,            
        top=2pt,              
        bottom=2pt]
        {
        \textbf{Prompt:} The user has visited categories \{\blue{(Gym, Mon, 5pm), (Restaurants, Mon, 6pm)}\}. Now is \{\blue{Mon}\} at \{\blue{8pm}\}, based on the user's categorical transition preference and categorical temporal preference, predict users' next most likely visiting `Category'. (format: Category)
    
        \textbf{Output:} \emph{\textbf{Bars.}}
    
        \textbf{Prompt:}
        The user actually visited \{\blue{Department} \blue{Store}\}.
        Based on this, what is the new insight you can get for the user's categorical transition preference? Generate the updated categorical transition preference. (format:\{category-category,...\})
    
        \textbf{Output:} \emph{\textbf{\{Restaurants-Bars, Pet Services-Restaurants, Restaurants-Hotel, Gym-Restaurants, Restaurants-Department Store\}}.}}
    \end{tcolorbox}

    \begin{tcolorbox}[title={Stage 3: Neighbor Preference Summarization},
        boxsep=2pt,           
        left=2pt,             
        right=2pt,            
        top=2pt,              
        bottom=2pt]
        {
        \textbf{Prompt:} The users' geographical neighbors' categorical transition preferences are \{\blue{Subway-Gym, Coffee Shop-Gym}\}; the users' semantic neighbors' categorical transition preferences are \{\blue{Restaurants-Bar, Bar-Movie Theater}\}; the users' social neighbors' categorical transition preferences are \{\blue{Restaurants-Plaza, Department Store-Clothing}\}. Summarize the neighbors' categorical transition preferences by considering their commonalities. (format: \{category-category,...\}).
    
        \textbf{Output:} \textit{\textbf{\{Gym-Subway, Gym-Coffee Shop, Restaurants-Bar, Bar-Movie Theater, Train Station-Plaza, Department Store-Restaurants\}}}}
    \end{tcolorbox}

    \begin{tcolorbox}[title={Stage 4: Next POI Recommendation},
        boxsep=2pt,           
        left=2pt,             
        right=2pt,            
        top=2pt,              
        bottom=2pt]
        {
        \textbf{Prompt:} Now is \{\blue{Mon}\} at \{\blue{9am}\}, based on the users’ current category sequence \{\blue{(Department Store, Mon, 8am), (Coffee Shop, Mon, 8am)}\}, his own categorical transition preference and categorical temporal preference, and his neighbors’ categorical transition preference and categorical temporal preference, predict the user’s next most likely visiting ‘category’. (format: category)
        
        \textbf{Output:} \textit{\textbf{Gym}}
    
        \textlangle Next Region and Distance Prediction\textrangle   
        
        \textbf{Prompt:} 
        Candidate POIs: (format:(Name, Category, Region, Distance)): \{\blue{(Gold's Gym, Gym, r2, 12km), (Anytime Fitness, Gym, r1, 5km), (CVS Pharmacy, Health, r1, 2km), ...}\}
        Given users' current check-in sequence \{\blue{(River 127, Hotel, r1, Mon, 8am), (Espresso Bar, Coffee Shop, r1, Mon, 8am)}\}, 
        recommend 10 POIs from the candidate set, considering his next likely visiting category \{\blue{Gym}\}, region\{\blue{r1}\}, and distance\{\blue{2km}\}. 
        State the reason for each recommendation and rank the importance of category, region, and distance preferences. \\
        (format: \{POI: reason; [importance ranking])\}}
    
        \textbf{Output:}
        \textbf{\textit{\{Anytime Fitness: The user’s next likely category is Gym. Anytime Fitness is a Gym, aligns with this preference. Anytime Fitness is in Region R1, matching the user’s preferred region. While 5km is not as close as 2km, it is still the best option among the gyms available. The CVS Pharmacy, while closer, does not match the Gym category; [category, region, distance].\}}}
    \end{tcolorbox}

\section{Conclusion}\label{sec:conclusion}

In this paper, we propose the novel MRP-LLM, which is tailored to perform a more accurate next POI recommendation and preserve check-in privacy with LLMs. 
Specifically, it first sets multiple subtasks and leverages a self-reflection mechanism to iteratively extract and refine users' fine-grained POI-related preferences. During recommendation, it retrieves neighbor users' signals to facilitate the next POI recommendation. Besides, all sensitive data are kept on local devices and transmitted with differential privacy. Extensive experiments on three real-world datasets demonstrate that MRP-LLM could effectively enhance LLMs' performance on the next POI recommendation task and achieve the best privacy-utility balance.



\clearpage
\newpage
\clearpage
\bibliographystyle{IEEEtran}
\bibliography{ref}

\end{document}